# Design Issues for providing Minimum Rate Guarantees to the ATM Unspecified Bit Rate Service

Rohit Goyal, Raj Jain, Sonia Fahmy, Bobby Vandalore, Shivkumar Kalyanaraman
Department of Computer and Information Science, The Ohio State University
395 Dreese Lab, 2015 Neil Avenue
Columbus, OH 43210-1277, U.S.A.
E-mail: {*goyal,jain*}@cis.ohio-state.edu

**Abstract**

*We describe proposed enhancements to the ATM Unspecified Bit Rate (UBR) service that guarantee a minimum rate at the frame level to the UBR VCs. These enhancements have been called Guaranteed Frame Rate (GFR). In this paper, we discuss the motivation, design and implementation issues for GFR. We present the design of buffer management and policing mechanisms to implement GFR. We study the effects of policing, per-VC buffer allocation, and per-VC queuing on providing GFR to TCP/IP traffic. We conclude that when the entire link capacity is allocate to GFR traffic, per-VC scheduling is necessary to provide minimum throughput guarantees to TCP traffic. We examine the role of frame tagging in the presence of scheduling and buffer management for providing minumum rate guarantees.*

## 1 Introduction

The ATM Unspecified Bit Rate (UBR) service category is expected to be a popular service for a variety of internetworking applications. In particular, TCP/IP based applications will extensively use the ATM UBR service category. ATM standards do not specify any congestion control mechanisms for the basic UBR service [2]. Switches are allowed to drop UBR cells when their buffers overflow. The absence of network based congestion control can lead to poor end to end performance for UBR based applications. As a result, competitive UBR implementations are expected to enhance the vanilla UBR service with intelligent tagging and buffer management policies. Per-VC based buffer management policies can significantly improve TCP performance over UBR [8]. Providing a guaranteed minimum rate to VCs has also been suggested as a means of improving TCP performance over UBR.

Guaranteed Frame Rate has been proposed in the ATM Forum to ensure minimum rate guarantees to UBR virtual circuits (VCs). The rate guarantee is provided at the frame level. GFR also guarantees the ability to share any excess capacity fairly among the GFR VCs. In this paper, we describe the various options available to the network to provide minimum rate guarantees to the UBR traffic class. We discuss the GFR proposals and outline the basic characteristics of GFR.

We explore three different mechanisms – policing, buffer management, and scheduling – for providing minimum guarantees to TCP traffic. We conclude that per-VC scheduling (fair queuing) is necessary to ensure minimum rate guarantees at the TCP level. We also discuss the dynamics of the interactions of per-VC scheduling with buffer management and policing. In this paper, we present simulation results for infinite TCP traffic.

## 2 Previous Work: TCP/IP over UBR

We have studied TCP performance over UBR for a variety of network latencies. In our studies, we have used an N-source symmetrical TCP configuration with unidirectional TCP sources. This configuration is illustrated in figure 1. Our experiments have included low latency configurations as well as large latency satellite configurations. The effect of higher priority VBR traffic has also been studied. The following subsections briefly summarize our previous results. Details about these experiments can be found in [8, 9, 10].

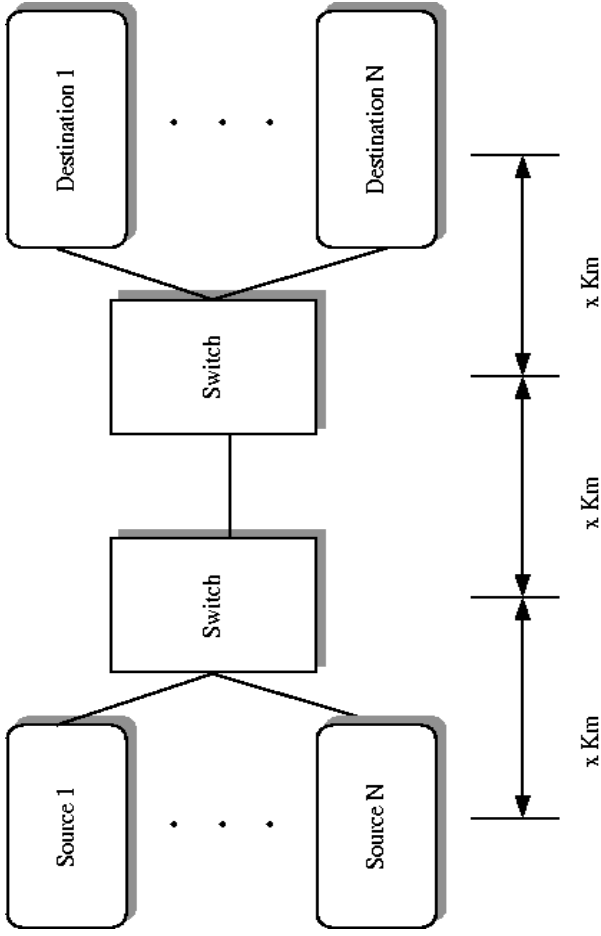

Figure 1: The N source TCP configuration

The performance of TCP over UBR is measured by the efficiency and fairness metrics which are defined as follows:

Efficiency = (Sum of TCP throughputs)/(Maximum possible TCP throughput)

The TCP throughputs are measured at the destination TCP layers. Throughput is defined as the total number of bytes delivered to the destination application, divided by the total simulation time. The results are reported in Mbps.

The maximum possible TCP throughput is the throughput attainable by the TCP layer running over UBR on a 155.52 Mbps link. For example, for a TCP segment size of 1024 bytes, the ATM layer receives 1024 bytes of data + 20 bytes of TCP header + 20 bytes of IP header + 8 bytes of LLC header + 8 bytes of AAL5 trailer. These are padded to produce 23 ATM cells. Thus, each TCP segment results in 1219 bytes at the ATM Layer. From this, the maximum possible throughput = 1024/1219 = 84% = 127.8 Mbps approximately on a 155.52 Mbps link (149.7 Mbps after SONET overhead).

Fairness Index = $(\Sigma x_i)^2 / (n \times \Sigma x_i^2)$

Where $x_i$ = throughput of the $i$th TCP source, and $n$ is the number of TCP sources. The fairness index metric applies well to our N-source symmetrical configuration.

In most cases, the performance of TCP over vanilla UBR is poor. The performance can be improved with enhanced drop policies and end system policies. A summary of our previous results is presented below:

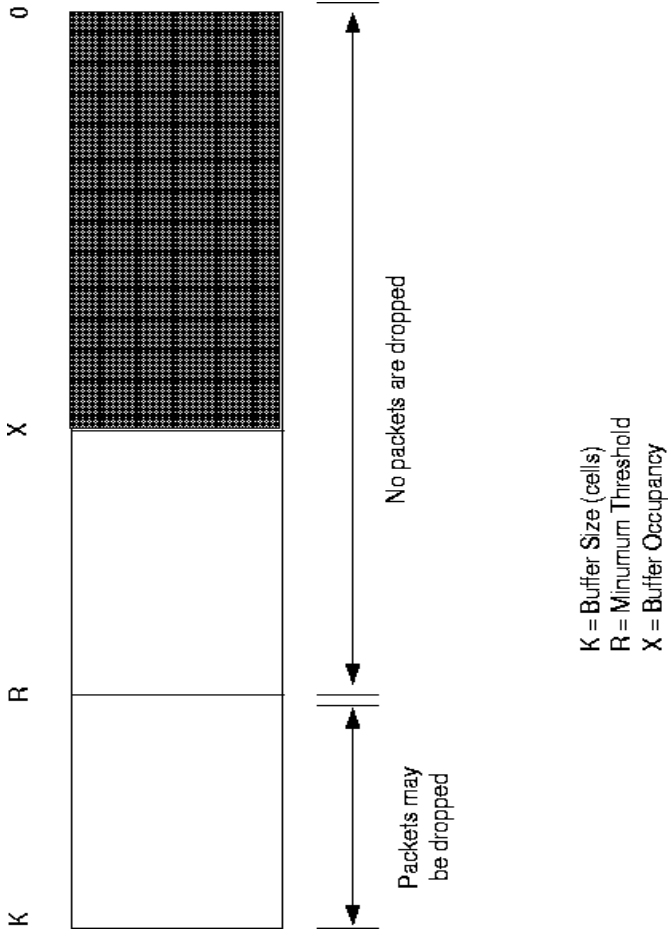

Figure 2: Drop behavior of buffer management policy

- TCP achieves maximum possible throughput when no segments are lost. To achieve zero loss for TCP over UBR, switches need buffers equal to the sum of the receiver windows of all the TCP connections.
- With limited buffer sizes, TCP performs poorly over vanilla UBR switches. TCP throughput is low, and there is unfairness among the connections. The coarse granularity TCP timer is an important reason for low TCP throughput.
- Efficiency typically increases with increasing buffer size.
- Fast retransmit and recovery improve performance for low latency configurations, but degrade performance in long latency configurations.
- SACK TCP (TCP with selective acknowledgements) improves performance especially for large latency networks.
- Early Packet Discard improves efficiency but not fairness.
- Per-VC buffer management improves both efficiency and fairness.

## 2.1 Presence of higher priority VBR traffic

When higher priority VBR traffic is present in the network, TCP over UBR may get considerably lower link capacity than without VBR. Moreover, the presence of VBR traffic could result in the starvation of UBR traffic for periods of time during which VBR uses up the entire link capacity. When VBR has strict priority over UBR, TCP (over UBR) traffic is transmitted in bursts and the round trip time estimates for the TCP connection become highly variable. An underestimation of the RTT is likely to cause a false timeout in the TCP indicating congestion even though the TCP packet is queued behind a VBR burst. An overestimtion of the RTT may result in much time being wasted waiting for a timeout when a packet is dropped due to congestion.

In this situation, we found that a minimun bandwidth guarantee improves TCP performance over UBR when the TCP connection may be starved for periods longer than the round trip propagation delay. A constant amount of bandwidth provided by a guaranteed rate ensures that TCP keeps receiving ACKs from the destination. This reduces the variation in the round trip times. Consequently, TCP is less likely to timeout. Simulation results with TCP over UBR with guaranteed rate can be found in [10].

# 3 Guaranteed Frame Rate

## 3.1 Introduction

Guaranteed Frame Rate (GFR) has been recently proposed in the ATM Forum as an enhancement to the UBR service category. It is as yet undecided if GFR will be a service within UBR or will be a separate service category. Guaranteed Frame Rate is intended to provide a minimum rate guarantee to VCs at the frame level. In addition, the GFR service also recommends the fair usage of any unused network capacity. GFR requires minimum signaling and connection management functions, and depends on the network's ability to provide a minimum rate to each UBR VC. GFR is likely to be used by applications that can neither specify the sustainable cell rate and burst size parameters needed for a VBR VC, nor can be subject to the ABR source rules (for rate based feedback control). Current internetworking applications fall into this category, and are not designed to run over QoS based networks. These applications could benefit from a minimum rate guarantee by the network, along with an opportunity to fairly use any additional bandwidth left over from higher priority connections. In the case of heterogenous internetworks, network elements outside the ATM network could also benefit from GFR guarantees. For example, IP routers separated by an ATM network could use GFR VCs to exchange control messages.

## 3.2 GFR Service Specification

The GFR specification provides a conformance definition for the frames of GFR VCs. The specification is based on a leaky bucket mechanism that identifies conforming and non-conforming frames. The frame level GFR guarantees must be translated into cell level guarantees for ATM. The minimum guaranteed frame rate is translated into a Minimum Cell Rate (MCR) parameter. Thus, MCR corresponds to the cell rate (in cells/sec) corresponding to the frame rate for frames of at most the maximum frame size (CPCS-SDU size). The maximum frame size is translated into a Maximum Burst Size (MBS) parameter where MBS is set to *two times the maximum frame size in cells.* This ensures the proper borrowing of tokens when the last and first cells of two packets arrive back to back. Since guarantees are made at the frame level, the ATM layer must also have information about the beginning and end of frames. In the case of AAL5 frames, the last cell of the frame has a flag marking the end of the frame. This cell is called the End of Message (EOM) cell.

The MCR and MBS are used as parameters to a leaky bucket mechanism (also known as the Generic Cell Rate Algorithm (GCRA) in ATM terminology) that determines the frames that are conforming to the minimum frame rate, and should be guaranteed service at that rate. Figure 3 illustrates the leaky bucket mechanism used to specify conforming frames. The leaky bucket consists of tokens that arrive into the bucket at the rate of MCR. When the first cell of a frame arrives at the network, if the number of tokens in the bucket is at least MBS/2 (i.e., there are at least one frame worth of tokens in the bucket), then all the cells of that frame are considered conforming cells, otherwise, all cells of that frame are non-conforming cells. Each conforming cell consumes a token from the leaky bucket. Non-conforming cells do not consume tokens from the bucket. Note that in figure 3, the Peak Cell Rate (PCR) refers to the maximum rate at which cells can arrive or leave the network element.

Non-conforming frames may be tagged or dropped by the network. Tagging is performed by setting the Cell Loss Priority (CLP) bit in the cell headers. The above policing mechanism allows tagging of complete frames. In general, tagging may be done by the end systems (hosts) or at the entrance to the ATM network. End systems could tag frames that are of lower priority. In this way, tagging can be used as an indication to the network about the relative priority of frames in a connection. On the other hand, frames that are non-conforming to the leaky bucket specification could be tagged at the entrance to the ATM network. This form of tagging can provide conformance information to the ATM network. When the network becomes congested, it can drop tagged frames in preference to untagged frames. In ATM terminology, the stream of untagged cells (whose CLP bit is zero) of a connection is called the CLP0 stream, while the entire cell stream of a connection is called the CLP0+1 stream.

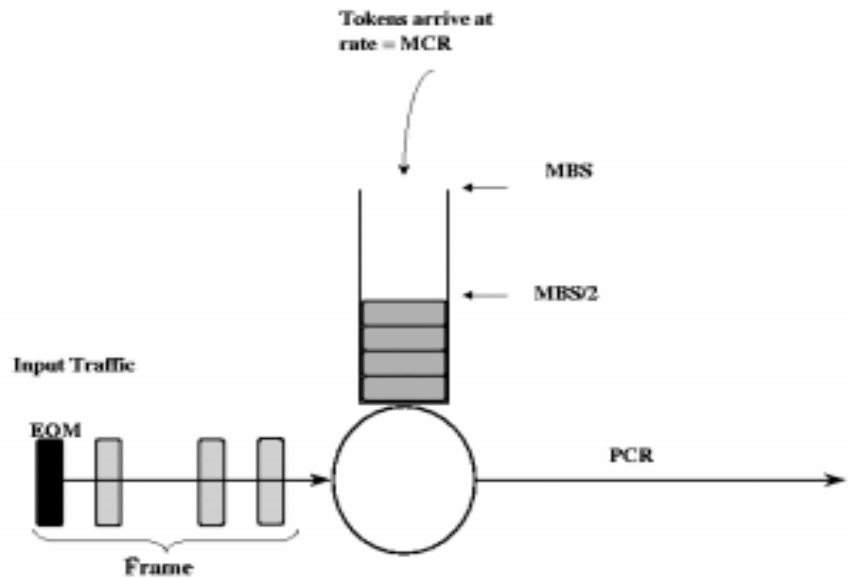

Figure 3: Leaky Bucket Policer

If a VC's entire traffic flow is conforming to the policing mechanism, then the entire flow should be serviced at MCR. If some frames of the flow are non-conforming, then the network should still provide service at MCR. In addition, the network should allocate to the VC, a fair share of any unused bandwidth, and in case of congestion, should drop tagged frames in preference to untagged frames. Note that *semantic information* about frame priority can only be determined at the end systems. Network tagging only determines the *conformance* of the frames. If the network and the end system both perform tagging, then the network tagging destroys the semantic information provided by the end system tagging. As a result, the network may choose not to tag the non-conforming frames, and use end system tagging information by the source to preferentially drop tagged packets. However, the network could tag all frames of some VCs to introduce a dual priority among the VCs. For example, IP routers at the edge of an ATM network may tag packets of all flows except those that belong to the network manager's flow.

# 4 Design Options for Implementing Minimum Rate Guarantees

There are three basic design options that can be used by the *network* to provide the per-VC minimum rate guarantees similar to GFR – policing, buffer management, and queueing. We briefly describe the options below, and in the following subsections, we discuss the mechanisms for each of the three options and describe our implementations of each. We then present our simulation results with various combinations of policing, buffer management and queuing to provide minimum rate guarantees. In our discussion, we use GFR as a model of a service that provides minimum rate guarantees. In this paper, we assume that a connection admission control (CAC) policy is present that ensures that minimum rate guarantees can indeed be provided to the active VCs.

The three mechanisms for providing per-VC minimum rate guarantees are:

1. **Tagging (policing):** *Network based tagging* can be used as a means of marking non-conforming packets before they enter the network. This form of tagging is usually performed when the connection enters the network. Figure 4 shows the role of network based tagging in providing a minimum rate service in a network. Network based tagging on a per-VC level requires some per-VC state information to be maintained by the network and increases the complexity of the network element. Tagging can isolate conforming and non-conforming traffic of each VC so that other rate enforcing mechanisms can use this information to schedule the conforming traffic in preference to non-conforming traffic. In a more general sense, policing can be used to discard non-conforming packets, thus allowing only conforming packets to enter the network. A sample tagging implementation based on the GFR conformance is discussed in section 4.1.

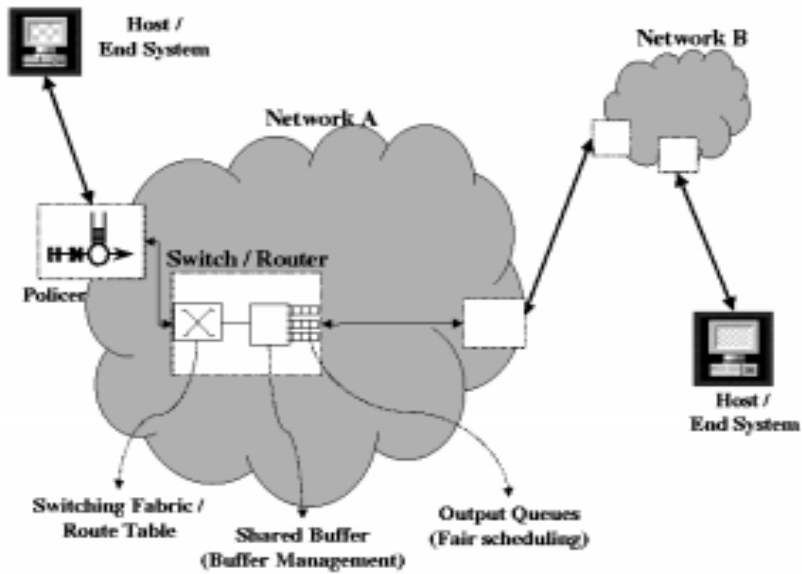

Figure 4: Network Architecture with policing, buffer management and scheduling

2. **Buffer management:** Buffer management is typically performed by a network element (like a switch or a router) to control the number of packets entering its buffers. In a shared buffer environment, where multiple VCs share common buffer space, per-VC buffer management can control the buffer occupancies of individual VCs. Per-VC buffer management uses per-VC accounting to keep track of the buffer occupancies of each VC. Figure 4 shows the role of buffer management in the connection path. Examples of per-VC buffer management schemes are Selective Drop and Fair Buffer Allocation [8]. Per-VC accounting introduces overhead, but without per-VC accounting it is difficult to control the buffer occupancies of individual VCs (unless non-conforming packets are dropped at the entrance to the network by the policer). Section 4.2 discusses a per-VC accounting based buffer management scheme for GFR.

3. **Queuing:** Figure 4 illustrates the position of queuing in providing rate guarantees. While tagging and buffer management control the entry of packets into a network element, queuing strategies determine how packets are scheduled onto the next hop. FIFO queuing cannot isolate packets from various VCs at their egress. As a result, in a FIFO queue, packets are scheduled in the order in which they enter the buffer. Per-VC queuing, on the other hand, maintains a separate queue for each VC in the buffer. A scheduling mechanism can select between the queues at each scheduling time. The role of queuing in minimum rate guarantees is further discussed in section 4.3.

## 4.1 Network Based Tagging

In GFR, network based tagging is used to identify the conforming packets. Conformance to the GFR service is specified by a generic cell rate algorithm (GCRA) or leaky bucket based mechanism as described in section 3.2. The conformance definition maps the frame level GFR guarantees to cell level guarantees.

The corresponding GCRA algorithm is outlined below. This definition describes a continuous state version of the leaky bucket algorithm [2]. The GCRA parameters are the Interval $I = 1/MCR$, and the Limit $L = CDVT + BT/2$ where $BT = (MBS - 1)(1/MCR - 1/PCR)$. Here BT is the burst tolerance corresponding to the maximum burst size (MBS). This specifies the token bucket depth in time units. CDVT is the cell delay variation tolerance associated with the rate. In the case of GFR, the tolerance in MCR is minimal, and the CDVT values is expected to be small. MBS is defined as twice the maximum frame size that can be sent by the application using the GFR connection.

Let the first cell of a frame arrive at time $t$. Let $X$ be the value of the leaky bucket counter , and $LCT$ be the last compliance time of a cell. Initially $X = LCT = 0$.

When the first cell of a frame arrives (at time $t$):

```
X1 := X - (t - LCT)
IF (X1 < 0) THEN
        X1 := 0
        CELL IS CONFORMING
        TAGGING := OFF
ELSE IF (X1 > BT/2 + CDVT) THEN
        CELL IS NON-CONFORMING
        TAG CELL
        TAGGING := ON
ELSE
        CELL IS CONFORMING
        TAGGING := OFF
ENDIF

IF (CELL IS CONFORMING)
        X := X1 + I
        LCT := t
ENDIF
```

When subsequent cells of a frame arrive (at time $t$) then:

```
IF (TAGGING == ON) THEN
        TAG CELL
ELSE
        CELL IS CONFORMING
        X1 := MAX(X - (t - LCT), 0)
        X  := X1 + I
        LCT = t
ENDIF
```

An exception can be made for the last cell of the frame (e.g., EOM cell for AAL5 frames). Since the last cell carries frame boundary information, it is recommended that the last cell should not be dropped unless the entire frame is dropped. For this reason, a network may choose to not tag the last cell of a frame. In our implementation we do not drop or tag the last cell.

When a network element sees a tagged frame, it cannot tell if the frame was tagged by the source end system or an intermediate network. This has significant influence in providing rate guarantees to the CLP0 stream, and is further discussed in section 4.2.

## 4.2 Buffer Management

Various buffer management schemes can be used as mechanisms for congestion avoidance and control. These include preferential dropping of tagged frames over untagged frames when mild congestion is experienced, and the use of per-VC accounting to fairly allocate buffers among the competing connections.

The original GFR proposal [11] outlined two possible techniques for providing minimum rate guarantees to UBR VCs:

- Using per-VC queuing (fair scheduling) and per-VC accounting (buffer management).
- Using FIFO queuing and per-VC policing (GCRA based tagging/dropping).

It was shown by [3] that for TCP traffic, it is difficult to provide end to end rate guarantees with tagging and FIFO queuing. A combination of large frame size and a low buffer threshold (at which tagged frames are dropped) is needed to provide minimum rate guarantees. This is undesirable because low thresholds result in poor network utilization.

It is also clear that per-VC queuing together with per-VC buffer management can provide minimum rate guarantees. However, it was unclear at that point, if the addition of per-VC buffer management to FIFO queuing with per-VC policing (tagging) would be enough provide the necessary rate guarantees. Also, with per-VC queuing, the role of tagging needs to be studied carefully.

We designed a buffer allocation policy called **Weighted Buffer Allocation (WBA)** based on the policies outlined in [8]. This policy assigns weights to the individual VCs based on their MCRs. The weights represent the proportion of the buffer that the connection can use before its frames are subject to discard. The WBA policy is described below:

Figure 2 illustrates the conditions under which frames may be dropped due to buffer overflow. When a the first cell of a frame arrives on a VC, if the current buffer occupancy (X) is less than a threshold (R), then the cell and the remaining cells of the frame are accepted. If the buffer occupancy exceeds this congestion threshold, then the cell is subject to drop depending on two factors – is the cell tagged? and, is the buffer occupancy of that VC more than its fair share?

Under mild congestion conditions, if there are too few cells of a VC in the buffer, then an attempt is made to accept conforming cells of that VC. As a result, if the VC's buffer occupancy is below its fair share, then a conforming (untagged) cell and the remaining cells of the frame are accepted, but a tagged cell and the remaining cells of the frame (except the last) are dropped. In this way, at least a minimun number of untagged frames are accepted from the VC.

If the buffer occupancy exceeds the mild congestion threshold, and the VC has at least its share of cells in the buffer, then the VC must be allowed to fairly use the remaining unused buffer space. This is accomplished in a similar manner as the schemes in [8], so that the excess buffer space is divided up equally amongst the competing connections.

The switch output port consists of a FIFO buffer for the GFR class with the following attributes:

- **K:** Buffer size in cells.
- **R:** Congestion threshold in cells ($0 \leq R \leq K$).
- **X:** Number of cells in the buffer.
- **Yi:** Total number of cells of VCi in the buffer.
- **Li:** Total number of untagged cells of VCi in the buffer.
- **Z:** Parameter ($0 \leq Z \leq 1$)
- **Wi:** Weight of VCi (for example Wi = MCRi/(Total UBR capacity)). $\Sigma Wi < 1$
- **Na:** Number of active VCs, i.e, VCs with cells in the buffer.

When the first cell of a frame arrives:

```
IF (X < R) THEN
        ACCEPT CELL AND FRAME
ELSE IF (X > R) THEN
        IF ((Li < R*Wi) AND (Cell NOT Tagged)) THEN
                ACCEPT CELL AND FRAME
        ELSE IF ((Yi - R*Wi)*Na < Z*(X-R)) THEN
                ACCEPT CELL AND FRAME
        ELSE
                DROP CELL AND FRAME (except the EOM cell)
        ENDIF
ENDIF
```

Further drop policies like EPD can also be used to drop all frames in excess of a more severe congestion threshold on top of the WBA threshold R. In our simulations we do not use EPD with WBA.

Per-VC buffer management can be effectively used to enforce tagging mechanisms. If the network does not perform tagging of VCs, but it uses the tagging information in the cells for traffic management, then per-VC accounting of untagged cells becomes necessary for providing minimum guarantees to the untagged (CLP0) stream. This is because, if a few VCs send excess traffic that is untagged, and no per-VC accounting is performed, then these VCs will have an excess number of untagged cells than their fair share in the buffer. This will result in conforming VCs (that tag all their excess traffic) receiving less througput than their fair share

The conforming VCs will experience reduced throughput because the non-conforming but untagged cells of the overloading VCs are not being isolated. This is the case even if per-VC queuing in a *shared buffer* is performed. If all VCs share the buffer space, then without per-VC buffer management, the overloading VCs will occupy an unfair amount of the buffer space, and may cause some conforming frames of other VCs to be dropped. Thus, no matter how frames are scheduled (per-VC or FIFO), if the switch allows an an unfair distribution of buffer occupancy, then the resulting output will also be unfair.

If a network does not tag the GFR flows, its switches should not use the information provided by the CLP bit unless they implement per-VC buffer management. We will show in section 6 that with only per-VC queuing and EPD (without per-VC buffer management or policing), a switch can provide rate guarantees for the CLP0+1 flow, but not for the CLP0 flow.

### 4.3 Fair Queuing (per-VC scheduling)

Fair queuing can control the outgoing frame rate of individual VCs. It was shown by [3] that with network tagging, fair queuing with preferential dropping of tagged frames (without per-VC management) can provide end to end rate guarantees for infinite TCP traffic. Our simulation results verify the intuition that that minimum rate guarantees for the CLP0+1 flow can be provided simply by fair queuing and EPD like buffer management. There is no need to look at the CLP bit to provide rate guarantees to the CLP0+1 flow. If the source end system does not tag its excess traffic, then it is arguable if the CLP0 flow has any special meaning for the end system. In this case, it might be just as appropriate in terms of end to end throughput to provide a GFR guarantee to the CLP0+1 flow.

To provide rate guarantees to the CLP0 flow, per-VC accounting of the CLP0 flow must be performed along with per-VC scheduling. This protects the network from overloading sources that do not tag their excess traffic.

We implement a Weighted Fair Queuing like service discipline to provide the appropriate scheduling. The details of the discipline will be discussed in a later paper. The results of our simulations are discussed in section 6.

## 5 Simulation of SACK TCP over GFR

This section presents the configuration for simulations of the various GFR mechanisms presented in the previous sections.

## 5.1 The Simulation Model

All simulations use the N source configuration shown in figure 1. All sources are identical and infinite TCP sources. The TCP layer always sends a segment as long as it is permitted by the TCP window. Moreover, traffic is unidirectional so that only the sources send data. The destinations only send ACKs. The delayed acknowledgement timer is deactivated, and the receiver sends an ACK as soon as it receives a segment. The version of TCP used is SACK-TCP. SACK TCP uses selective acknowledgements to selectively retransmit lost segments. SACK-TCP has been shown to improve the performance of TCP especially for large delay networks. Details about our implementation of SACK TCP can be found in [9].

Most of our simulations use 15 sources. We also perform some simulations with 50 sources to understand the behavior with a large number of sources.

Link delays are 5 milleseconds. This results in a round trip propagation delay of 30 milliseconds. The TCP segment size is set to 1024 bytes. For this configuration, the TCP default window of 64K bytes is not sufficient to achieve 100% utilization. We thus use the window scaling option to specify a maximum window size of 600,000 Bytes.

All link bandwidths are 155.52 Mbps, and peak cell rate at the ATM layer is 149.7 Mbps after the SONET overhead. The duration of the simulation is 20 seconds. This allows for adequate round trips for the simulation to give stable results.

In our simulations, the end systems do not perform any tagging of the outgoing traffic. We measure performance by the end to end effective throughput obtained by the destination TCP layer. The goal is to obtain an end-to-end througput as close to the allocations as possible. Thus, in this paper we look at the aggregrate CLP0+1 flow.

# 6 Simulation Results

## 6.1 N TCP sources with equal rate allocation

We simulated N TCP sources on a 155 Mbps link. Each source was allocated 1/Nth of the link capacity. We used only a per-VC accounting based buffer management policy called Selective Drop [8]. Per-VC queuing or policing were not used. Figure 5 shows the resulting end to end TCP performance with switch buffer sizes 12000 cells and 24000 cells for 15 and 50 sources.

Figure 5 plots the average efficiencies of the 4 simulation experiments. For the 15 (50) source configurations, each source has a maximum expected throughput of about 8.2 (2.5) Mbps. The actual achieved throughput is divided by 8.2 (2.5) , and then the mean for all 15 (50) sources is found. The point plotted represents this mean. The vertical lines show the standard deviation below and above the mean. Thus, shorter standard deviation lines mean that the actual throughputs were close to the mean. Standard deviation is an indicator of the fairness of the sources' throughputs. A smaller standard deviation indicates larger fairness.

In figure 5, all TCP throughputs are within 20% of the mean. The mean TCP throughputs in the 15 source configuration are about 8.2 Mpbs for both buffer sizes. For 50 sources, the mean TCP throughputs are about 2.5 Mbps. The standard deviation from the mean increases with increasing buffer sizes for the 15 source configuration. The increase in buffer sizes leads to more variability in the allocation, but as the number of sources increases, the allocation reduces, and so does the variability. All configurations exhibit almost 100% efficiency, and the fairness for large number of sources is high. From this we conclude that *equal allocation of the available bandwidth for TCP traffic to the CLP0+1 flow can be achieved by per-VC accounting only.*

## 6.2 N TCP sources with unequal rate allocations

The N TCP connections were divided up into 5 categories of bandwidth allocations. For example, 15 sources are divided into groups of 3, with MCRs of approximately 2.6 Mbps, 5.3 Mpbs, 8 Mpbs, 10.7 Mbps, 13.5 Mpbs.

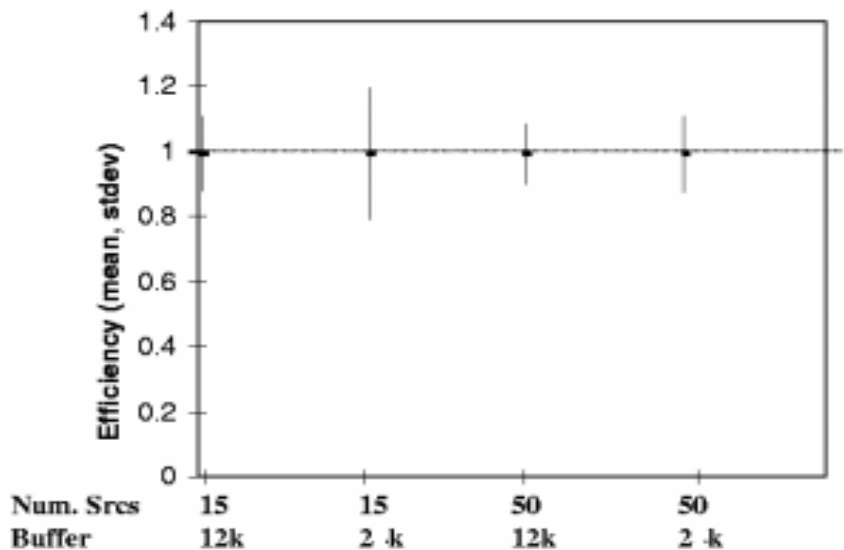

Figure 5: Per-VC Accounting: Equal Rate Allocations

To achieve unequal allocations of end to end throughputs, we tried two options:

1. Per-VC accounting (Weighted Buffer Allocation) with tagging.

2. Per-VC scheduling without tagging or accounting.

Figure 6 shows the resulting allocations obtained from the simulations for tagging and per-VC accounting. The 5 points represent the efficiencies obtained by the 5 categories of bandwidth allocations. The horizontal dotted line represents the target efficiency (100%) desired from the allocations. Thus, the average allocation of the first category was over 2.5 times its target allocation (2.6 Mbps), while the last category received 0.8 of its allocation (13.5). The dotted line represents the target efficiency for each category. The total efficiency (not shown in the figure) was almost 100%. The figure shows that per-VC accounting with tagging cannot effectively isolate competing TCP traffic streams. We tried various drop thresholds R for WBA, and R = 0.5 achieved the best isolation. However, the achieved rates were not close to the target rates. *FIFO buffering with Per-VC buffer allocation together with tagging is not sufficient to provide GFR guarantees.* This conclusion is also verified by [4] for a different per-VC buffer allocation policy.

Figure 7 shows the resulting allocations obtained for the 15 source unequal rate configuration with per-VC scheduling without tagging or accounting. The figure shows that the achieved throughputs are close (within 10%) to the target throughputs for the CLP0+1 stream. Also, the total efficiency (not shown in the figure) is over 90% of the link capacity. From this, we conclude that *per-VC queuing alone is sufficient to provide end to end minimum rate guarantees for infinite TCP traffic.*

To provide rate guarantees to the CLP0 flow, per-VC accounting is needed to separate the CLP0 frames of the different VCs. If tagging is performed by the network on all the VCs, then per-VC accounting may not be necessary, and preferential dropping of tagged frames with per-VC scheduling should be sufficient. Simulation results of this configuration are under current study.

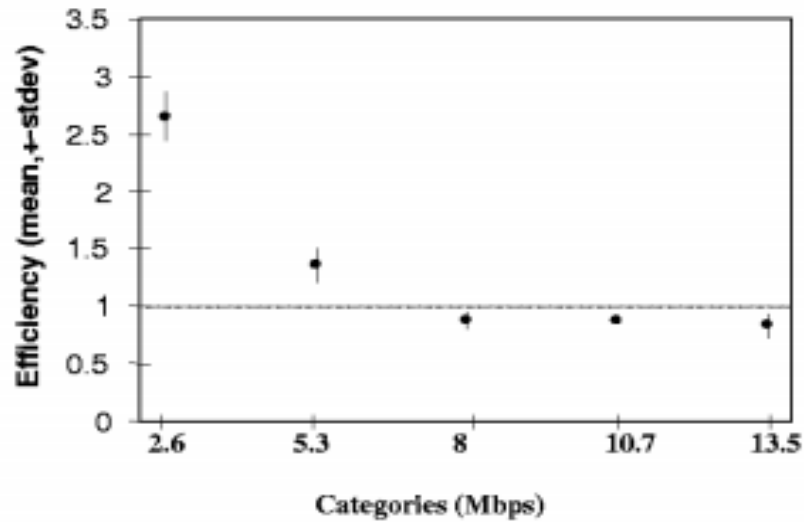

Figure 6: Per-VC Accounting+Tagging: Unequal Rate Allocations

# 7 Summary

To summarize, our goal has been to find ways to provide GFR guarantees to infinite TCP traffic flows. So far, we have considered the case where the TCP end systems do not perform tagging. We measure performance by the end to end effective TCP throughputs of the CLP0+1 stream. The following conclusions can be drawn for infinite TCP/IP traffic.

- FIFO queuing with per-VC accounting based buffer management is not sufficient to provide GFR guarantees.
- GFR guarantees to the CLP0+1 stream can be provided by per-VC queuing (fair scheduling) alone. This is irrespective of any tagging operations.
- To provide guarantees to the CLP0 flow, the network must perform per-VC buffer management in addition to per-VC queuing. This protects the CLP0 traffic of conforming flows from flows whose non-conforming frames are not tagged either by the end system or by the network.

The results so far are summarized by table 1. The table shows the eight possible options available with per-VC accounting, tagging (by the network) and queuing. An "X" indicates that the option is used, and a "-" indicates that it is not used. The GFR column indicates if GFR guarantees can be provided by the combination specified in the row.

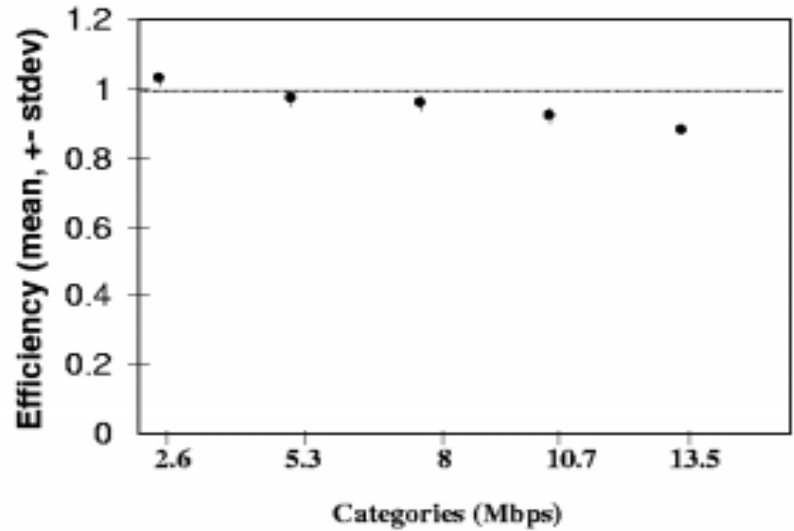

Figure 7: Per-VC Scheduling: Unequal Rate Allocations

Table 1: Design options for providing GFR guarantees

| Per-VC Accounting | Per-VC Tagging | Per-VC queuing | GFR | Notes |
|---|---|---|---|---|
| - | - | - | No | Clearly |
| X | - | - | No | Only for equal allocations |
| - | X | - | No | Shown by [3] |
| X | X | - | No | Shown here |
| - | - | X | Yes | Only for the CLP0+1 stream |
| X | - | X | Yes | |
| - | X | X | Yes | Results under current study |
| X | X | X | Yes | Also for CLP0 stream |

[4] Debashis Basak, Surya Pappu, "GFR Implementation Alternatives with Fair Buffer Allocation Schemes," ATM FORUM 97-0528, July 1997.

[5] John B. Kenney, "Satisfying UBR+ Requirements via a New VBR Conformance Definition," ATM FORUM 97-0185.

[6] John B. Kenney, "Open Issues on GFR Frame Discard and Frame Tagging," ATM FORUM 97-0385.

[7] M. Mathis, J. Madhavi, S. Floyd, A. Romanow, "TCP Selective Acknowledgement Options," Internet RFC 2018, October 1996.

[8] R. Goyal, R. Jain, S. Kalyanaraman, S. Fahmy and Seong-Cheol Kim, "UBR+: Improving Performance of TCP over ATM-UBR Service," Proc. ICC'97, June 1997.

[9] R. Goyal, R. Jain et.al., "Selective Acknowledgements and UBR+ Drop Policies to Improve TCP/UBR Performance over Terrestrial and Satellite Networks," To appear, Proc. IC3N'97, September 1997. [1]

---

[1] All our papers and ATM Forum contributions are available from http://www.cis.ohio-state.edu/~jain